\documentstyle[aps,prl,epsf,twocolumn]{revtex}
\begin{document}
\title {
``Intermittency" in Hydrodynamic Turbulence \\
      as an Intermediate Asymptotic to Kolmogorov'41 Scaling}
\author{Victor S. L'vov and Itamar Procaccia}
\address{
Departments of Physics of Complex Systems and Chemical Physics,
Weizmann Institute of Science, Rehovot, 76100, Israel}
\maketitle
\begin{abstract}
A synopsis of an analytical theory of scaling in developed turbulence
is proposed on the basis of the Navier-Stokes equations.  It is shown
that corrections to the normal Kolmogorov 1941 scaling behavior of
the $n$-th order velocity structure functions are finite Re effects
which disappear when the inertial interval exceeds 5-6 decades. These
corrections originate from  the correlation between the  velocity
differences and energy dissipation which are characterized by an
anomalous (subcritical) exponent.  The values of the experimentally
observed scaling indices for the $n$-th order structure functions for
$n$ between 4 and 14 are in agreement with our findings.
\end{abstract}
\pacs{PACS numbers 47.10, 47.25C}
The desire to find a universal description of turbulence that was
initiated by Richardson\cite{R926} in the 1920's, seemed for a while
to be satisfied by the bold suggestion of Kolmogorov\cite{K941} in
1941 (K41) that one may construct a theory with one
universal scaling exponent.  This exponent was ascribed to
differences of the longitudinal velocity across a
scale $R$,
$\delta u_\ell ({\bf x}+{\bf R},{\bf x}) $ $\equiv [{\bf u} ({\bf
x}+{\bf R})- {\bf u} ({\bf x})]\cdot {\bf R}/R$  $\sim R^{1/3}$, in the
sense that the structure functions $S_n(R)$ of $\delta u_\ell$ satisfy the
scaling laws
\begin{equation}
S_n(R)\equiv \langle\lbrack\delta
u_\ell ({\bf x}+{\bf R},{\bf x})\rbrack ^n \rangle \sim (\bar\varepsilon
R)^{n\zeta _n}\sim (\bar\varepsilon R)^{n/3}
\label{A1}
\end{equation}
for values of $R$ in the ``inertial range" $L\gg R\gg\eta$ where $L$ and
$\eta$ are the integral scale and the Kolmogorov dissipation scale
respectively. In (\ref{A1}) $\langle \dots \rangle$ denotes an average
over time, and $\bar\varepsilon$ is the mean of the dissipation field
$\varepsilon({\bf x},t)$ $\equiv  \nu \lbrack\partial_\alpha u_\beta({\bf
x},t) +\partial_\beta u_\alpha({\bf x},t) \rbrack^2/2$ ($\nu$ is the
kinematic viscosity).  This suggestion of Kolmogorov was immediately
attacked both  by theorists (e.g. Landau) and experimentalists.  Indeed,
it is rather astonishing that a problem like fluid turbulence, which
suffers from very large fluctuations and strong correlations, should be
amenable to such a simple description; even Kolmogorov himself revised his
thinking\cite{K962} and changed (\ref{A1}) to a more complicated form
(which fell under attack as well).  One measurement that raised a lot of
objections to the K41 approach is the measurement of the correlation
function of the dissipation field $K_{\varepsilon\varepsilon}(R)$
\begin{equation}
K_{\varepsilon\varepsilon}(R)=\langle\hat\varepsilon
({\bf x}+{\bf R},t)\hat\varepsilon ({\bf x},t)\rangle
\propto R^{-\mu}\,,
\label{A2}
\end{equation}
where $\hat\varepsilon ({\bf x},t)=\varepsilon({\bf x},t)
-\bar\varepsilon$.  It was found that $K_{\varepsilon\varepsilon}(R)$
decays very slowly in the inertial range, with $\mu$ being in the range
0.2--0.4\cite{MS91}.  It was claimed that the K41 theory required $\mu$ to
vanish.  Accordingly, there have been many attempts to construct
models\cite{K962,NS64,M974,FSN8} of turbulence to take (\ref{A2}) into
account and to explain how the measured deviations $(\zeta_n-1/3)$ in the
exponents of the structure functions are related to $\mu$.

In this Letter we make use of a recent description due to L'vov and
Lebedev\cite{LL94a,LL94b} of the mechanism  for anomalous behavior of the
energy dissipation field. In their theory $\mu$ is an independent nonzero
scaling exponent while the  K41 scaling is exact in the limit Re$\to
\infty$.  We shall argue, in contradiction with the common wisdom, that in
the case $\mu=0$ we expect a breakdown of K41 scaling behavior.  For
positive values of $\mu$ the K41 scaling of $S_n(R)$ should be valid in
the limit of infinite Reynolds numbers Re.  However, the observed value of
$\mu$ is sufficiently small to be close to the borderline of the breakdown
of K41.  As a result there are essential corrections to (\ref{A1}) even at
the largest experimentally available values of Re which are $10^8-10^9$.
We shall estimate these corrections and show that the known observations
can be rationalized on the basis of a new picture of turbulence proposed
here.

Our starting point is the Navier-Stokes equations:
\begin{equation}
\partial{\bf u}/\partial t+({\bf u}\cdot\nabla ){\bf u}
-\nu\nabla^2{\bf u}-\nabla\,p=0\,,\quad \nabla\cdot{\bf u}=0 \ .
\label{A3}
\end{equation}
These can be used to derive equations
for $S_n(R,t)$: $\partial S_n(R,t)/\partial t+{\it D}_n(R,t)=
J_n(R,t)$, which in the stationary state yields the balance:\
$D_n(R)=J_n(R)$,
\begin{eqnarray}
D_n(R)&\equiv & n\langle\lbrack\delta u_\ell
({\bf x},{\bf x}')\rbrack ^{n-1}
\{ {\bf P}_\ell[{\bf u}({\bf x})\cdot\nabla~{\bf u}({\bf x})]
\label{A4}\\
&&-{\bf P}_\ell[{\bf u}({\bf x}')\cdot\nabla~{\bf u}({\bf x}')]
\} \rangle,
\nonumber \\
J_n(R)&=&\nu n\langle\lbrack \delta u_\ell ({\bf x},{\bf x}')\rbrack
^{n-1}\lbrack\nabla^2u_\ell({\bf x})-\nabla^{'2}u_\ell({\bf
x}')\rbrack \rangle,
\label{A5}
\end{eqnarray}
and $R=|{\bf x}-{\bf x}'|\,,~{\bf P}_\ell=\lbrace {\bf R}/R-\lbrack
({\bf R}/ R) \cdot \nabla^{-2}\nabla\rbrack\nabla\rbrace$.

In this Letter, rather than using formal diagrammatic expansions
\cite{LL94b} we follow  L'vov and Lebedev\cite{LL94a} in developing
physical reasoning to expose the physical mechanism for anomalous scaling
behavior.  We begin with $K_{\varepsilon\varepsilon}(R)$.  The reason for
the existence of dissipation correlations between two points separated a
large distance $R$ apart must be due to velocity components of size $R$
and smaller.  Eddies larger than $R$ only sweep the two points {\bf
x},{\bf x}' together, but cannot lead to correlations.  Consider the
simplest possible picture in which the velocity field ${\bf u}({\bf x},t)$
contains only  one typical scale $R$.  One may understand  this field as
the  result of filtering out (from a turbulent field with K41 spectrum)
all  the Fourier components lying in $k$ space out of some shell of radius
$k=2\pi /R$. We denote such a field as ${\bf V}_{_R} ({\bf  x},t)$, and we
can  estimate its gradient as ${\bf V}_{_R}/R\sim (\bar\varepsilon)^
{1/3}R^{-2/3}$. In terms of $S_2(R)$ the situation may be exemplified as
shown in Fig.1a.  We have a bell shaped contribution centered around $\log
R$, which in logarithmic scale has a width of $b/\Delta$.  This notation
will become clear later.  In such a {\sl $R$-band} flow the evaluation of
$K_{\varepsilon\varepsilon}(R)$ is $\nu^2 (V_{_{R}}/R)^4$ $\sim \nu^2
(\bar \varepsilon)^{4/3}R^{-8/3}$.  Consider next a {\sl two-band} flow in
which there are two scales of eddies, $R$ and $r$, with $R\gg r\gg\eta$.
The structure function has now two bell shaped contributions, both of
logarithmic width $b/\Delta$.  In evaluating $K_{\varepsilon
\varepsilon}(R)$ we shall interpret the ensemble average as a two step
process. The first one, $\left\langle \dots \right\rangle_{{r,}_R}$ is  a
conditional average on an ensemble of $r$-eddies at a given velocity field
${\bf V}_{_R} ({\bf x},t)$.  The second, $\left\langle \dots
\right\rangle_{_R}$ is an average over an ensemble of $R$-eddies:
\begin{eqnarray}
\langle\hat\varepsilon({\bf
x}_1,t)\hat\varepsilon ({\bf x}_2,t) \rangle&=&
\left\langle\left\langle\varepsilon ({\bf x}_1,t)
\hat\varepsilon ({\bf x}_2,t)\right\rangle_{{r,}_R}\right\rangle_{_R}
\nonumber\\
&=&\left\langle\left\langle\hat\varepsilon ({\bf   x}_1,t)
\right\rangle_{{r,}_R}\right\langle\hat\varepsilon ({\bf x}_2,t)
\rangle_{{r,}_R}\rangle_{_R}.
\label{A6}
\end{eqnarray}
The last step is justified because the dissipative field is mostly
sensitive to the fast, small scale motion; each $r$-ensemble average can
be done in the presence of some realization ${\bf V}_{_R} ({\bf x},t)$ of
the $R$-eddies, and only when we compute the correlations we need to
average in the $R$-ensemble.  Consider now the average over small eddies
$\langle\hat\varepsilon ({\bf x}_1,t)\rangle_{{r,}_R}$.   It has a
contribution from $|\nabla {\bf V}_{_R}({\bf x},t)|^2$ which is almost not
affected by the averaging over the small scales, and then it has
contributions from the small eddies.  The largest contribution to
$\varepsilon ({\bf x}_1,t)$ comes from $|\nabla {\bf V}_r({\bf x},t)|^2$
which is even larger than $|\nabla {\bf V}_{_R}({\bf x},t)|^2$ (${\bf
V}_r({\bf x},t)$ denotes the velocity of $r$-scale eddies). However this
contribution falls off when we subtract $\bar\varepsilon $ to compute
$\hat\varepsilon ({\bf x}_1,t)$.  Next order contributions come from the
effect of $\nabla {\bf V}_{_R}$ on the distribution function of
$\varepsilon$-fluctuations.  We can therefore expand
$\langle\hat\varepsilon ({\bf x}_1,t)\rangle_r/\nu \approx$
$
\approx |\nabla{\bf V}_{_R}|^2
+a_{\alpha\beta}\partial_\alpha (V_{_R})_\beta
+b |\nabla{\bf V}_{_R}|^2+c|\nabla{\bf V}_{_R}|^3+\dots
$
where we have displayed the tensor indices in the second term and
suppressed them in the rest.  The coefficients in the expansion are
computed in the $r$-ensemble which is isotropic and homogeneous, and
therefore $a_{\alpha\beta}$ $\propto \delta_{\alpha\beta}$.  Thus the
second term vanishes by incompressibility.  It is clear that the
coefficient $b$ is dimensionless, and that $c$ has the dimension of
time.  Since the coefficients $b$ and $c$ are characteristic of the
$r$-ensemble only, we can estimate them again in a K41 manner.  Thus
$b\sim(\bar\varepsilon r)^0,~c\sim\bar\varepsilon^{1/3}r^{-2/3}$.  In
other words, $b$ is a constant, independent of the scale $r$, whereas
$c$ is of the order of the turnover time $\tau_r$.  The ratio of the
fourth to the third terms is a ratio of two time scales, $c|\nabla
{\bf V}_{_R}|\sim \tau_r/\tau_{_R}$ which is small if $r \ll R$.  We
therefore write finally
\begin{equation}
\langle\hat\varepsilon({\bf x}_1,t)\rangle_{{r,}_R}
\approx\nu |\nabla{\bf V}_{_R}({\bf x}_1,t)|^2 (1+b)\ .
\label{A7}
\end{equation}
The central point of the argument is the fact that $b$ is independent
of the scale $r$.  It implies that $\langle\hat\varepsilon ({\bf
x}_1,t)\rangle_r$ is renormalized in the {\sl same way by every
$r$-eddy contribution}. Therefore when we go to the full
spectrum, see Fig. 1b, we need to multiply our answer $b$ by the
number of statistically independent channels of interaction.  This
number is estimated as $(\log R-\log \eta)/(b/\Delta)$.  In total,
\begin{equation}
\langle \hat \varepsilon ({\bf x}_1,t)\rangle_{{r,}_R}
\approx \nu |\nabla{\bf V}_{_R}({\bf x}_1,t)|^2\lbrack (1+\Delta\,
\log(R/\eta)\rbrack.
\label{A8}
\end{equation}
Notice that the width $b/\Delta$ is endowed now with a physical meaning.
It measures the characteristic length in logarithmic scales of the
statistical dependence of the turbulent motion.  Next consider the
contribution of three groups of motions on scales $R$, $x$ and $y$, with
$R\gg x\gg y\gg\eta$.  We discussed already the contribution of the direct
interaction between the scales $R$ and $x$ and $R$ and $y$. Now we wish to
evaluate the indirect two-step effect on the dissipation field due to
effects of the largest $R$ motion on the smallest $y$-motions via the
intermediate $x$-motions.  By repeating the above arguments we find
instead of (\ref{A7}) $\langle\langle \hat\varepsilon ({\bf
x}_1,t)\rangle_y\rangle_x $ $\approx \nu |\nabla{\bf
V}_{_R}|^2(1+b_x+b_y+b_xb_y+...)$ where we skip for brevity the additional
indices in $\langle\langle\dots\rangle_{y,x}\rangle_{x,_R}$ which denote
the conditions of averaging. Subscripts on the $b$'s remind us of their
origin.  In fact they are all the same, independent of scale.  The number
 of contributions proportional to $b^2$ would have been $[(\log
R-\log\eta)/(b/\Delta)]^2$ if the relation between $x$ and $y$ were
arbitrary.  Since they are ordered in size $x<y$ we have only a half of
that number.  Finally $\langle\langle\hat\varepsilon ({\bf
x}_1,t)\rangle_y\rangle_x\approx$
$
\nu |\nabla{\bf V}_{_R}|^2(1+\Delta~\log(R/\eta)+\case{1}{2}[
\Delta~\log(R/\eta)]^2\ .
$
Going on to $n$-step interactions we will find additional
contributions $[\Delta\log(R/\eta)]^n/n!$, where the $n!$ arises
again due to the scale ordering $x> y> z\dots$  Such a series
resums to
\begin{equation}
\langle\langle\langle\hat\varepsilon ({\bf x}_1,t)\rangle _\eta
\rangle _y...\rangle\approx \nu |\nabla{\bf V}_{_R}({\bf x}_1,t)|^2
({R/\eta})^\Delta\ .
\label{A9}
\end{equation}
Using this result in the calculation of
$K_{\varepsilon\varepsilon}(r)$ we find
\begin{equation}
K_{\varepsilon\varepsilon}(R)\propto R^{-8/3+2\Delta}\ .
\label{A10}
\end{equation}
This result was first obtained in\cite{LL94a,LL94b}, and in the context of
a passive scalar in\cite{LPF4}, and in both places it was supported by
fully resummed diagrammatic expansions.  Together with (\ref{A2}), Eq.
(\ref{A10}) implies $\Delta=4/3-\mu/2\approx 1.2$.  The physical meaning
of (\ref{A9}) is that $({R/\eta})^\Delta$ is simply the total number of
effective channels of interaction.

Armed with this understanding we can return now to analyze the
balance equation, following the approach of\cite{LPF4,K994}.  We
start with $J_n$, Eq. (\ref{A5}).  Adding ${\bf r}$ to ${\bf x}$ and
${\bf x}^{\prime}$, replacing $\nabla_x$ and $\nabla_{x'}$ with
$\nabla_r$, and integrating once by parts we find
\begin{eqnarray}
J_n(R)&=&-4\nu n(n-1) \langle\lbrack\delta
u_\ell ({\bf x},{\bf x}')\rbrack^{n-2}
\Xi ({\bf x},{\bf x}')\rangle\,,
\label{A11} \\
\Xi ({\bf x},{\bf x}')&\equiv &
\lbrack | \nabla u_\ell ({\bf x}) |^2
-\nabla~u_\ell ({\bf x})\cdot\nabla  u_\ell ({\bf x}')\rbrack  \ .
\nonumber
\end{eqnarray}
Consider first the ``decoupled" part of $J_n(R)$ with respect to $\Xi
({\bf x},{\bf x}')$. Up to correction of order $(\eta /R)^{4/3}\ll
1$, $\langle\Xi ({\bf x},{\bf x}')\rangle =\bar \varepsilon /3$ and
\begin{equation}
J_n^{\rm dec}(R)= -\case{4}{3}n(n-1)\bar\varepsilon S_{n-2}(R)\ .
\label{A12}
\end{equation}
Next, in the ``coupled"  contribution to (\ref{A11}) the term $|\nabla u_\ell
({\bf x})|^2$ contributes $\varepsilon ({\bf x})/3$, and the term $\nabla
u_\ell ({\bf x})\cdot\nabla 'u_\ell ({\bf x}')$ gives a negligible
contribution.  The  reason follows from the previous arguments.  Repeating
the multistep averaging process discussed above over the motions of
intermediate scales from $\eta$ to $R$, we get in each step a partly
isotropized contribution for the next step.  This multistep process will
enhance the spherically symmetric part of the quantities averaged upon.  The
quantity $|\nabla u_\ell ({\bf x})|^2$, which is the gradient of the
longitudinal part, gives, after averaging over angles, just a third of
$\varepsilon ({\bf x})$.  In contrast, a $\nabla u_\ell ({\bf x})$ term, as a
vector, will contribute very little.  A more detailed analysis shows that the
vector terms are smaller by a factor of $(\eta /R)$. Therefore
\begin{eqnarray}
J_n^{\rm c}(R)&\approx&  -\case{4}{3}n(n-1)
\langle\lbrack\delta u_\ell ({\bf x},{\bf x}')\rbrack^{n-2}
\lbrack \varepsilon ({\bf x})-  \bar \varepsilon \rbrack  \rangle
\label{A13} \\
\approx \nu n(n & - & 1)\big[ C^{\rm red}_n S_n^{\rm red}(R)
+ C^{\rm irr}_n S_n^{\rm irr}(R)\big]
R^{\Delta-2}/\eta^\Delta \  .
\nonumber
\end{eqnarray}
The last line follows from (\ref{A9}) and from  estimating $|\nabla{\bf
V}_{_R}|^2$ as  $\delta u({\bf x},{\bf x}')^2/R^2$.  In (\ref{A13}) $C_n^{\rm
red}$ and $C_n^{\rm irr}$ are  dimensionless coefficients, $C_2^{\rm
red}=C_2^{\rm irr}=0$.  $S_n^{\rm red}(R)$ is the sum of all contributions to
$S_n(R)$ which are obtained from decoupling the average $\langle \delta
u_\ell ({\bf x},{\bf x}')^n \rangle$ into factors like $\langle \delta u_\ell
({\bf x},{\bf x}')^m \rangle\langle \delta u_\ell ({\bf x},{\bf x}')^{n-m}
\rangle$, excluding the contributions taken in $J_n^{\rm dec}$.  $S_n^{\rm
irr}(R)$ is the remaining part that cannot be decoupled.  Note that ``{\sl
coupled}"  and ``{\sl decoupled}" contributions are also ``{\sl reducible}"
and ``{\sl irreducible}", but in a special sense, i.e.  with respect to
$|\nabla{\bf V}_{_R}|^2$ which originated from $\varepsilon$.

$\!$To compare $J_n^{\rm c}$ with $J_n^{\rm dec}$ we can
evaluate them using the
K41 estimate $S_n(R)\!\sim\! (\bar \varepsilon R)^{n/3}$.
The ratio $J_n^{\rm c}/J_n^{\rm dec}\!\!\sim$ $(\eta / R)^{4/3\!
-\!\Delta}\!\!
=\!\!(\eta/R)^{\mu/2}$.  We thus see that when $\!\Delta$ reaches a critical
value of $\case{4}{3}$, or equivalently when the exponent $\mu$ becomes zero,
$J_n^{\rm c}$ is of the same importance as $J_n^{\rm dec}$.  Since
experimentally $\mu$ is not zero, but quite small, $\mu\!\approx\! 0.3$, we
expect that $J_n^{\rm c}$ will introduce via the balance equation a
visible correction to the scaling exponents even for very large Re.
We shall estimate these corrections next.

We need first to estimate $D_n(R)$.  As we discussed before, one
gradient in a correlation is not likely to build up a multistep
telescopic contribution.  It can be substantiated that ${\it D}_n(R)$
can be estimated order by order in perturbation theory by local
integrals\cite{LPF4}.  We divide ${\it D}_n(R)$ again into coupled
and decoupled  contributions in the same sense as in $J_n(R)$, i.e.
with respect to $\delta u({\bf x},{\bf x}')^{n-2}$ on the one hand
and the rest on the other.  There are $n-1$ possibilities to do that.
The resulting estimate is
\begin{eqnarray}
D_n^{\rm dec}(r)&=&\case{1}{2} n(n-1) D_2(R)S_{n-2}(R)\,,
\label{A14}\\
D_n^{\rm c}(r)&\approx&n \big[ d^{\rm red}_n S_{n+1}^{\rm red}(R)
+ d^{\rm irr}_n S_{n+1}^{\rm irr}(R)\big] / R\ .
\label{A15}
\end{eqnarray}
Here $d^{\rm red}_n$ and $d^{\rm irr}_n$ are  dimensionless
coefficients, $d^{\rm red}_2=0$.  For $n=2$ (\ref{A12}--\ref{A15})
give  $D _2(R)\sim S_3(R)/R$, $J_2(R)\approx - 8\bar\varepsilon/3$. The
balance equation yields $S_3(R)/R \sim \bar\varepsilon$.  We find no
correction to $S_3(R)$, as is expected by the requirement that
$S_3(R)=-{4\over 5}\varepsilon R+6\nu dS_2(R)/dR$.  Note that
(\ref{A15}) is valid only for $n\geq 2$, so we do not generate
corrections to $S_2$ in our approach either. In order to find corrections
for $n\geq 4$ consider the balance equation $D_n(R)=J_n(R)$  for
$n\geq 3$. Equations (\ref{A12}, \ref{A14}) show that the decoupled
contributions on both sides are identical. Thus $D_n^{\rm
c}(R)=J_n^{\rm c}(R)$. Together with (\ref{A13},\ref{A15}) it gives
\begin{eqnarray}
&&\big[ d^{\rm red}_n S_{n+1}^{\rm red}(R)
+ d^{\rm irr}_n S_{n+1}^{\rm irr}(R)\big]
\label{A16}  \\
&\approx& \big(\bar \varepsilon R \big)^
{\case{1}{3}}\big(\eta/R\big)^{\mu/2}
\big[ C^{\rm red}_n S_n^{\rm red}(R)
+ C^{\rm irr}_n S_n^{\rm irr}(R)\big]
\nonumber
\end{eqnarray}
The RHS of  (\ref{A16})   vanishes in the limit $\eta
/R\rightarrow 0$. Therefore  $S_{n+1}^{\rm irr} \sim S_{n+1}^{\rm
red}\propto (\bar \varepsilon R)^{n/3}$. This means that in the limit
Re$\to \infty$ we recover the K41 scaling  of the structure
functions.  For nonzero $\eta /R$ one may consider  (\ref{A16}) as a
recurrence relation which expresses $S^{\rm irr}_{n+1}$ in terms of
lower order structure functions $S_m$, $m\leq n$. Considering in
(\ref{A16}) the case $n=3$ and taking into account the fact that
$S_3(R)\sim \bar \varepsilon R$ we find the first nontrivial
 correction in   $S_4^{\rm irr}$, namely $\sim \nu S_3(R)R^{\Delta-2
}/\mu^\Delta$.  Finally
\FL
\begin{equation}
S_4(R)=S^{\rm red }_4+S^{\rm irr}_4 \sim
(\bar\varepsilon R)^{4/ 3}\Big[1 + \tilde C_4 \Big({\eta
\over R}\Big)^{\mu/ 2}\Big]\,,
\label{A17}
\end{equation}
with  $\tilde C_4 \simeq C_3^{\rm irr}/d_4^{\rm red}$.  We stress that the
value $\mu =0$ corresponds to non-decaying correlations of the dissipation as
a function $R$. In that case the physics of turbulence should change
completely. For the hypothetical case $\mu   <  0$ we expect the multiscaling
behavior to exist even in the limit Re$\to \infty$. There is an indication of
such a possibility  in the problem of turbulent diffusion of a passive scalar
field\cite{LPF4}.  In this case, with $\mu> 0$, the correction term vanishes
when $\eta/R\to 0$.  Notwithstanding, due to the smallness  of $\mu$, for any
appreciable value of $\tilde C_4$ we need an enormously large inertial range
before $ \tilde C_4 (\eta / R)^{\mu/2}$ becomes negligible compared to unity.
The experimental evidence\cite{OP94,LP94} is that the coefficients involved
are larger than $7$. For $\mu\simeq 0.3$ the correction term is actually {\sl
dominant} as long as $\eta / R > 10^{-5}$. At the present values of
experimentally available Re there is no case in which there exist 5 orders of
magnitude of inertial range. It would thus appear that $S_4(R)$ scales nicely
with an apparent exponent whose value is $4/3-\mu/2$.

The comparison of our theoretical considerations with experiments
becomes more difficult for $n>4$, since the data analysis in all
experiments to date did not take into account the contributions of the
reducible parts to $S_n$. The experimental data analysis is
equivalent to assuming that the irreducible parts of $S_n$ are always
much larger than the reducible ones. If we adopt this assumption, the
recurrence relation (\ref{A16}) would result in the prediction
\begin{equation}
n\zeta_n=n/3 - \mu (n-3)/2\ .
\label{A18}
\end{equation}
This prediction  can be compared with experiments.
In recent experiment\cite{SSJ3} the values of $n\zeta_n$ shown in
Table 1 were reported.  Choosing $\mu$=0.276 we get the theoretical
values shown in the Table.  For comparison the K41 values are also
displayed.  We stress however that the excellent agreement between
theory and experiment cannot be taken too seriously due to the
unavoidable existence of reducible contributions. It just shows
that there are no glaring contradictions.

Equation (\ref{A18}) has a superficial similarity to the
$\beta$-model, in which $n\zeta_n=n/3 - \mu (n-3)/3$.  A major
difference is that in the $\beta$-model this correction is believed
to be asymptotic also at Re$\rightarrow\infty$, and that here we have
completely different physics, in which (\ref{A18}) is an intermediate
asymptotic result for large but finite  Re.  Also, our theory does
not exclude the possibility that the existence of non-negligible
contributions with K41 scaling may influence the apparent scaling
exponents in a way that results in a non-linear dependence $n\zeta_n$
vs.  $n$. The theoretical message is that the K41 scaling for the
structure functions remains exact in the limit Re$\rightarrow\infty$,
and probably leaves non negligible contributions at experimentally
relevant values of Re.  We propose that the experimental data
analysis should be redone in the light of this theory to separate the
contributions with the K41 exponents and those with anomalous
exponents. In doing so one has to be also aware of other possible
corrections to K41 which stem from  noncritical mechanisms, like the
corrections $\delta\zeta_2$ $\sim(\eta/L)^{2/5}$ which are due to the
anisotropy of the excitation of turbulence on the outer scale
$L$\cite{LPb4,GLLP}. Such corrections may lead to differences in the
measured exponents in different experiments.

In summary we have presented a new theory of scaling behavior in
turbulence at finite Re. In contrast to various phenomenological
models of intermittency our approach is based on the Navier-Stokes
equations. The first cornerstone in this theory is the fact that the
K41 scaling of velocity differences is exact in the limit Re$\to
\infty$; this is a non-perturbative result of the analytical
theory\cite{LL93}. The second cornerstone is the theoretical
understanding of how fields that are sensitive to the dissipative
scale, like the energy dissipation field, exhibit anomalous scaling
with nontrivial exponents\cite{LL94a,LL94b,LPF4}.  The last cornerstone is
the observation that the dissipation field is subcritical. Its scaling
exponent $\mu/2$ is very small:  $\mu/2\simeq 0.1-0.2$. We have shown here
that the Navier Stokes equations impose a constraint in the form of the
balance equation which feeds back the anomaly of the dissipative field
onto the scaling of the structure functions.  Since the exponent $\mu$ is
positive, the feed back effect disappears at infinite Re. Nevertheless,
due to   the smallness of $\mu/2$ the subcritical corrections to scaling
 remain important for all experimentally realizable Re. It is noteworthy
that our theory explains why the observed deviations from K41 are so
small:  there exists a small parameter in the theory, i.e.  $\mu/2$, which
allows us to find small corrections to K41.  This leaves the fascinating
theoretical problem of understanding whether the small value of $\mu$ is
accidental, or whether it stems from fundamental reasons.

$\!\!\!\!\!\!${\bf Acknowledgments.} Numerous discussions with
Volodya Lebedev were very instructive for us.  This work has been
supported in part by the Minerva Foundation, Munich, Germany and
the Basic Research Fund of the Israeli Academy.

\begin{figure}
\epsfxsize=8.7truecm
\epsfbox{fig1.eps}
\vskip.3truecm
\caption
{Different scale eddy contributions to the structure functions
(see text for details).}
\label{fig1}
\end{figure}

\begin{table}
\caption{Comparison between experimental and theoretical
values of the scaling exponents. Experiment $^{14}$: turbulent
bounadry layer at Re (based on its thickness) = 32000}%
{\begin{tabular} {|l||l|l|l|l|l|l|l|}
{}~~~~~~n & ~~2 & ~~4 & ~~6 & ~~8  & ~10 & ~12 &~14 \\ \hline\hline
$n\zeta_n$~(exp.)&0.70~~&1.20~~&1.62~~&2.00~~&2.36~~&2.68~~&3.02~~\\
\hline $n\zeta_n$~(theor.)&---&1.20&1.59&1.98&2.37&2.76&3.15\\ \hline
$n\zeta_n$~(K41)&0.67&1.33&2.00&2.67&3.33&4.00&4.67\\
\end{tabular}}%
\end{table}

\begin{references}
\bibitem{R926} L. Richardson, Proc. Roy. Soc. London, {\bf A110},
		709 (1926).
\bibitem{K941} A.N. Kolmogorov,
                Dokl. Akad. Nauk. USSR {\bf 38}, 538 (1941).
\bibitem{K962} A.N. Kolmogorov,
                J. Fluid Mech. {\bf 13}, 82 (1962).
\bibitem{MS91} K.R. Sreenivasan, C. Meneveau, Phys. Fluids {\bf
		A5}, 512 (1993).
\bibitem{NS64} E. Novikov, R Stewart, Izv. Geoph. {\bf 3},
		408 (1964).
\bibitem{M974} B.B. Mandelbrot,
      		J. Fluid Mech. {\bf 62}, 331 (1974).
\bibitem{FSN8} U. Frisch, P. Sulem and M. Nelkin,
		J. Fluid Mech. {\bf 87}, 719 (1978).
\bibitem{LL94a} V.S. L'vov, V.V. Lebedev,
                Europhys. Lett., submitted.
\bibitem{LL94b} V.V. Lebedev, V.S. L'vov,
                JETF Letters {\bf 59}, 577 (1994).
\bibitem{LPF4} V.S. L'vov, I. Procaccia  and A.L. Fairhall,
                Phys. Rev. E, in press.
\bibitem{K994} R.H. Kraichnan,
		Phys. Rev. Lett. {\bf 72}, 1016 (1994).
\bibitem{OP94}  A. Praskovsky, S. Oncley,
                Europhys. Lett., submitted.
\bibitem{LP94} V.S. L'vov, I. Procaccia,
                Europhys. Lett.,  submitted.
\bibitem{SSJ3} G. Stolovitzky, K.R.  Sreenivasan and A.  Juneja,
		Phys. Rev. E {\bf 48},   R3217 (1993).
\bibitem{LPb4} V.S. L'vov, I. Procaccia
                Phys. Rev. E, {\bf 49} 4044 (1994).
\bibitem{GLLP} S. Grossmann, D. Lohse, V.S. L'vov, I. Procaccia
                Phys. Rev. Lett.,  {\bf 73} 432 (1994).
\bibitem{LL93} V.V. Lebedev, V.S. L'vov,
                Phys. Rev. E {\bf 47}, 1794 (1993).
\end{references}
\end{document}